\documentclass[aps,prc,showpacs,twocolumn,preprintnumbers,floatfix,
showkeys,tightenlines]{revtex4}
\usepackage{amsmath,amssymb,amsfonts}
\usepackage{dcolumn}
\usepackage{graphicx}
\usepackage{color}
\allowdisplaybreaks
\begin{document} 

\title{Symmetry energy of warm nuclear systems}

\author{B. K. Agrawal$^{1}$, J. N. De$^{1}$, 
 S. K. Samaddar$^{1}$, M. Centelles$^{2}$ 
 and X. Vi\~nas$^{2}$} 
\affiliation{
$^1$Saha Institute of Nuclear Physics, 1/AF Bidhannagar, Kolkata
{\sl 700064}, India \\
$^2$Departament d'Estructura i Constituents de la Mat\`eria,
Facultat de F\'{\i}sica, \\
and Institut de Ci\`encies del Cosmos, Universitat de Barcelona, \\
Diagonal {\sl 645}, {\sl 08028} Barcelona, Spain\\}


\begin{abstract}

The temperature dependence of the symmetry energy and symmetry 
free energy coefficients of infinite nuclear matter and of finite
nuclei is investigated. For infinite matter, both these coefficients
are found to have a weaker dependence on temperature at densities
close to saturation; at low but homogeneous densities, the temperature
dependence
becomes stronger. For finite systems, different definitions of
symmetry energy coefficients are encountered  in the 
literature yielding different  
values. A resolution to this problem is suggested from a
global liquid-drop-inspired fit of the energies and free energies
of a host of nuclei covering the entire periodic table. The hot nucleus
is modeled in a subtracted finite-temperature-Thomas-Fermi framework,
with dynamical surface phonon coupling to nucleonic motion plugged in.
Contrary to infinite nuclear matter, a substantial change in the
symmetry energy coefficients is observed for finite nuclei with 
temperature.

\end{abstract}



\maketitle

\section{Introduction}

The nuclear symmetry energy is a measure of the energy gain in converting
isospin asymmetric nuclear matter to a symmetric system.  Its value
depends  on  density and temperature.
It is assessed through the symmetry energy coefficient $e_{sym}(\rho ,T)$;
for an asymmetric system with density $\rho $, temperature $T$ and 
asymmetry $\delta $, the energy per nucleon, to a good approximation,
can be written as 
\begin{eqnarray}
e(\rho,\delta,T)=e(\rho,\delta =0,T)+e_{sym}(\rho ,T )\delta^2,
\end{eqnarray}
where 
$e_{sym}(\rho ,T)\delta^2 $ is the symmetry energy content per nucleon
of the system and $e(\rho, \delta =0,T)$ is the energy per nucleon
of symmetric nuclear matter.
 Here $\delta = (\rho_n-\rho_p)/(\rho_n+\rho_p) $
is the isospin asymmetry, $\rho_n$ and $\rho_p$ are the neutron and proton
densities of the system with $\rho =\rho_n+\rho_p $. The coefficient
$e_{sym}(\rho ,T)$ can then be defined, without loss of generality, as
\begin{eqnarray}
\label{eq:CE}
e_{sym}(\rho,T)= \frac{1}{2}\Bigl (\frac{\partial^2e(\rho,\delta,T)}{\partial
\delta^2} \Bigr )_{\delta=0}.
\end{eqnarray}

The density dependence of the symmetry energy is   instrumental
in understanding the behaviour of the equation of state (EOS) of
asymmetric nuclear matter.  Accurate knowledge of this  EOS  
is important in interpreting the structure
of finite nuclei away from the stability line, some critical issues in
astrophysics could only be addressed with this knowledge.  
Some progress has recently been made in understanding
the behaviour of the symmetry energy at the subnormal densities from
the analyses  of  the  isospin diffusion data in heavy ion collision
\cite{Tsang04,Chen05a,Li05a,Li08} and from the available data for
the neutron skin thickness of several nuclei \cite{Centelles09}.
The experimental data on the isotopic dependence of the nuclear giant
monopole resonance  in even-A Sn isotopes \cite{Garg07,Li07} also provide
some informations on the nuclear symmetry energy which are in agreement
with those derived from the analyses of the isospin diffusion data.
The behaviour of  the nuclear symmetry energy at supranormal densities is
largely unknown.  The precise measurements of the 
observable  properties of 
compact stars and the transport model analyses of 
heavy-ion collisions
at intermediate and high energies may provide some constraints on the
high density behaviour of the symmetry energy.

Understanding  the thermal behavior  of the symmetry energy is of
utmost contemporary importance.  It has a role in changing the nuclear
drip lines as the nuclei warm up. It is a key element in deciding a number
of serious issues in  the astrophysical scenario like supernova explosions
\cite{bar} or explosive nucleosynthesis. A large (small) value of the
symmetry coefficient, say, inhibits (accelerates)  the change of protons
to neutrons through electron capture \cite{ste,jan}.  The consequent
modulation of the EOS of hot nuclear matter through shift in isospin
asymmetry shapes the dynamical phases of the collapse and explosion 
of a
massive star. Furthermore, in this rapidly changing scenario, the hot and
possibly dilute nuclear matter is an inhomogeneous congregate of nuclear
clusters of different sizes. A precision understanding of the thermal
evolution of the symmetry coefficients of finite nuclei then becomes a
matter of seminal importance.  
Since the density
derivative $L$ of the symmetry coefficient $e_{sym}$
\begin{equation}
L=3\rho_0 \frac{\partial e_{sym}}{\partial \rho}\biggl |_{\rho_0}
\end{equation} 
reflects the pressure difference on the
neutrons and protons and is thus one of the determinants in fixing the
neutron skin  of nuclei \cite{hor}, 
the nature and stability of phases within a
warm neutron star, its crustal composition or its thickness \cite{ste1}
would be strongly influenced from the temperature dependence of symmetry
energy. The cooling rate of warmer proto-neutron stars through neutrino
convection \cite{rob} may also be affected from the thermal change 
of the symmetry energy.

A  few measurable bulk parameters often  help in giving a better  
feel on the characteristics of the nuclear EOS at zero temperature around
the saturation density.
At $T$=0, the first term in Eq.~(1) can be cast as
\begin{eqnarray}
e(\rho,\delta =0) \simeq a_v+\frac {1}{2}K_v\epsilon^2,
\end{eqnarray}
and the symmetry coefficient is written as
\begin{eqnarray}
e_{sym}(\rho) \simeq e_{sym}(\rho_0)-L\epsilon+\frac {1}{2} K_{sym}\epsilon^2,
\end{eqnarray}
where $\epsilon =(\rho_0-\rho )/3\rho_0 $, $a_v$ is the energy 
per nucleon of symmetric nuclear matter, 
$K_v$ is the isoscalar
density incompressibility and $K_{sym}$ is the symmetry incompressibility,
all at saturation density $\rho_0$.
The parameters $a_v,\rho_0, $ or $K_v$ are known more or less in
tight bounds \cite{dut},  while
$L$ and $K_{sym}$ are less precise
\cite{Li08,Centelles09,agr}. Mindful that the above characterization of the EOS
is  valid only close to the saturation density, the EOS of infinite
homogeneous nuclear matter spanning a wide density range has 
been calculated with an effective interaction as input, the interaction
designed so as to describe broadly the experimental observables 
quite well. The symmetry energies at different temperatures and densities 
have then been calculated \cite{xu,de,sam1,mou}. 

The above calculations for homogeneous nuclear matter have been done
in the mean-field (MF) approximation. Low density nuclear matter is,
however, not homogeneous. Formation of clusters of different sizes
\cite{fri,pei} becomes energetically more favorable there. 
A detailed knowledge
of the composition of nuclear matter is then needed to appreciate how the
energies and symmetry energies are affected because of clusterization. In
hot inhomogeneous supernova matter, the neutrino-driven energy transfer
\cite{jan} is dictated partly by the size of the clusters. Using the
virial expansion technique, clusterization in dilute nuclear matter and
its import in the evaluation of the symmetry energy and its temperature
dependence have been investigated by Horowitz and coworkers
\cite{hor1,con}, where they consider matter to be composed of only very
light particles whose scattering phase-shifts are known. The calculations
are later extended with inclusion of all possible heavy clusters by
exploiting the general analysis of the grand-canonical partition function
for nuclear matter in the $S-$matrix framework \cite{mal,de1,sam2}.
One interesting fallout of this calculation is that even symmetric nuclear
matter, when dilute and warm, may have sizeable symmetry energy content;
this is so because the produced clusters may not be all symmetric,
though the disassembled matter conserves 
overall  vanishing isospin asymmetry.
The persistence of this feature in clusterized dilute matter makes the
definition of the symmetry energy coefficient $e_{sym}$ 
or the symmetry free energy coefficient $f_{sym}$ ambiguous. The
symmetry coefficient $e_{sym}$ as defined by Eq.(\ref {eq:CE})  has been
approximated in the literature in several ways:
\begin{eqnarray}
e_{sym}(\rho,T)=[e(\rho,\delta,T)-e(\rho,\delta=0,T)]/\delta^2,
\end{eqnarray}
\begin{eqnarray}
e_{sym}(\rho,T)=e(\rho,\delta=1,T)-
e(\rho,\delta=0,T),
\end{eqnarray}
For homogeneous nuclear matter at densities upto saturation,
$e(\rho,\delta,T)$ is found to be linear in $\delta^2$ in the whole
range of $\delta $.
All the three definitions  (given by Eqs.~(2), (6) and (7)) 
yield nearly the same
value of $e_{sym}$. The total energy or free energy of the asymmetric
dilute inhomogeneous system, however, deviates strongly from
the above-mentioned linearity. This leads to different values \cite{de2}
for $e_{sym}$ or $f_{sym}$ from the three definitions. 
Definition given by Eq.~(7) has been used \cite{kow,nat}
in the experimental determination of the values of the symmetry 
coefficients of warm dilute nuclear matter.

For application in core-collapse supernova simulations, it is more
insightful to have understanding about the temperature dependence
of  the symmetry energy of isolated nuclei or nuclei embedded in a 
nucleonic medium \cite{de3}. In a limited temperature domain ($T \le $
2 MeV), calculations of the symmetry energy coefficients of atomic
nuclei have been attempted in a schematic model by Donati et al.,
\cite{don}. These calculations take into account the coupling of the 
nucleons to the dynamical surface phonons. This results in an increased 
effective nucleon mass, the so-called energy mass (or $\omega$-mass
$m_\omega $) coming into play. The $\omega$-mass 
carries  signatures of interaction that are nonlocal
in time \cite{has}. The energy mass decreases with temperature
\cite{pra}, this brings in a decreased density of states and thus an
increase in the symmetry coefficient. A fall-out is that electron
captures are hindered in supernova matter. From  shell-model
Monte-Carlo calculations in this limited temperature range, 
quantitative support to these findings were given by Dean et al.,
\cite{dea}. The symmetry coefficients so calculated are, however, 
much below the nominally acceptable range. Calculations have also been 
done recently by Lee and Mekjian \cite{lee} in a density functional 
theoretic approach, but the suitability of 
the approximations used
keeps the calculations trustworthy
only in a low temperature domain ($T \le $ 3 MeV).

Very recently, attempts have  been made 
to explore the thermal evolution
\cite{de4} of the symmetry energy coefficients of specific atomic
masses in an extended temperature region ($T \le $ 8 MeV) in the
finite-temperature Thomas-Fermi (FTTF) framework with subtraction
technique \cite{sur}. For a finite nucleus of mass $A$, the symmetry
coefficient is defined as in Eq.~(6) using the difference 
of the nuclear parts of the energy per nucleon of a nuclear
pair of mass  number $A$ having different isospins.
The definition from this difference method \cite{dea}, however, does
not prescribe a unique value of the symmetry coefficient for the 
nucleus; the values depend on the choice of the isospin
asymmetric isobar pair. Another definition \cite{sam1} based on the
local density approximation (LDA) has also been used. It has its drawback
too, it suffers from its dependence on the isospin of the nucleus
and also on the charge-dependent density distribution $\rho (r)$. 

An unambiguous
definition of the temperature dependent symmetry energies and the
symmetry coefficients can, however,  be obtained in the framework
of the liquid drop model. 
Very recently, the energies
and free energies of a large number of nuclei spanning almost
the entire periodic table were calculated \cite{de5} as a 
function of temperature in the subtracted FTTF method and fitted
in the spirit of the liquid drop model with volume, surface, Coulomb
and symmetry energy coefficients. The temperature-dependent coefficients
(including the symmetry coefficients) so obtained convey a sense 
of average, but they are ambiguity-free and in the present 
state of our knowledge serves the purpose.

The article is organized as follows. In Sec. II, we portray the
theoretical outlines. Sec. IIA discusses the  EOS of 
homogeneous nuclear matter. 
Hot nuclear systems are inherently unstable,
a suitable model for proffering stability to such systems 
is described in Sec. IIB; deriving relevant observables
from  such a description is also briefly outlined there.
A short discussion on the energy mass is given in Sec. IIC,
relevant formulas for calculating the symmetry coefficients are
presented in Sec IID.  
In Sec. III,
results and discussions are given. 
Sec. IV ends with the concluding remarks.

\section{Elements of theory}

The methodology employed to calculate the symmetry energy and symmetry
free energy of hot infinite and finite nuclear systems is described
in this section. Self-consistent thermal models are employed. 
Three interactions of the Skyrme class, namely, SkM$^*$ \cite{bar1},
SLy4 \cite{cha}, and SK255 \cite{agrawal03} are used to describe the systems.
Ground state bulk properties of nuclei over the whole periodic table
for $A \ge $16 are reproduced quite satisfactorily with these interactions.
Some bulk properties like the saturation density, symmetry energy coefficient,
isoscalar incompressibility, etc, for these interactions are given
in Table I.  The interaction SLy4 describes systems with extreme isospin
better, SK255 gives comparatively larger symmetry energy coefficient
and larger incompressibility. Different interactions are chosen to see
whether  different thermal sensitivity of the 
symmetry properties becomes manifest even if the ground-state energies
are described  similarly 
by these interactions.

\subsection{Energy density functionals }

 With the Skyrme energy density functional, the total energy density
of the system is written as,

\begin{eqnarray}
{\cal E }(r)&=&\frac {\hbar^2}{2m_{n,k}}\tau_n+\frac {\hbar^2}
{2m_{p,k}}\tau_p \nonumber \\
&&+\frac {1}{2}t_0[(1+\frac {1}{2}x_0)\rho^2-(x_0+\frac {1}{2})
(\rho_n^2+\rho_p^2)]  \nonumber \\
&&+\frac {1}{12}t_3\rho^\alpha [(1+\frac {x_3}{2})\rho^2
-(x_3+\frac {1}{2})(\rho_n^2+\rho_p^2)] \nonumber \\
&&+\frac {1}{16}[3t_1(1+\frac {1}{2}x_1)-t_2(1+\frac {1}{2}x_2)]
(\nabla \rho )^2 \nonumber \\
&&-\frac {1}{16}[3t_1(x_1+\frac {1}{2})+t_2(x_2+\frac {1}{2})]
[(\nabla \rho_n)^2+(\nabla \rho_p)^2] \nonumber \\
&&+{\cal E}_c(r).
\end{eqnarray}
In Eq.~(8), $t_0,t_1,t_2,t_3,x_0,x_1,x_2,x_3$ and $\alpha $ are the
Skyrme parameters (given in Table II for the interactions) and $\rho_n$
and $\rho_p$ are the neutron and proton densities ($\rho =\rho_n+\rho_p $).
For infinite homogeneous systems, the derivative terms do not contribute,
 neither  the Coulomb term ${\cal E}_c$, since the whole system is
charge neutral. The finite nucleus has contributions from the
Coulomb force though, as it is charged. The nucleon effective mass
$m_{q,k}$ is defined through
\begin{eqnarray}
\frac {m}{m_{q,k}(r)}&=&1+\frac {m}{2\hbar^2} \Bigl  \{[t_1(1+
\frac {x_1}{2})+t_2(1+\frac {x_2}{2})] \rho \nonumber \\
&&+[t_2(x_2+\frac {1}{2})-t_1(x_1+\frac {1}{2})]\rho_q \Bigr \}, 
\end{eqnarray}
with $q =(n,p)$ referring to neutrons or protons. This 
effective mass, often called the nucleon $k$-mass,
arises from the momentum dependence in the effective interaction. 
In the Thomas-Fermi approximation  at finite tempertature, 
the kinetic energy density is
\begin{eqnarray}
\tau_q=\frac {2m_{q,k}}{\hbar^2}A_q^*TJ_{3/2}(\eta_q),
\end{eqnarray}
with 
\begin{eqnarray}
A_q^*=\frac {1}{2\pi^2} \Bigl (\frac {2m_{q,k}T}{\hbar^2} \Bigr )^{3/2}.
\end{eqnarray}
The function $J_k(\eta )$ is the standard Fermi integral.
At zero temperature, $\tau_q$ reduces to the familiar expression,
\begin{eqnarray}
\tau_q=\frac {3}{5}(3\pi^2)^{2/3}\rho_q^{5/3}.
\end{eqnarray}
The fugacity $\eta_q$ is obtained as,
\begin{eqnarray}
\eta_q(r)=[\mu_q-V_q(r)]/T
\end{eqnarray}
where $V_q(r)$ is the single-particle potential experienced by
the nucleons (including the Coulomb part for the protons in finite
systems) and $\mu_q$ their chemical potentials.
The nucleonic density $\rho_q$ is related to $\eta_q$ by
\begin{eqnarray}
\rho_q(r)=A_q^*J_{1/2}(\eta_q(r)).
\end{eqnarray}
The Coulomb energy density is given by ${\cal E}_c(r)= {\cal E}_c^d(r)
+{\cal E}_c^{ex}(r)$, where the direct contribution is
\begin{eqnarray}
{\cal E}_c^d(r)=\frac {1}{2}\rho_p(r)\int \rho_p(r^\prime )\frac {
e^2}{|{\bf r} -{\bf r^\prime } |} d{\bf r^\prime },
\end{eqnarray}
and the exchange part, in the Slater approximation, is
\begin{eqnarray}
{\cal E}_c^{ex}=-\frac {3e^2}{4\pi}(3\pi^2)^{1/3}\rho_p^{4/3}(r).
\end{eqnarray}
In the Landau quasi-particle approximation, the entropy density
of the nucleons can be computed as 
\begin{eqnarray}
{\cal S}(r)=-\frac {2}{h^3}\sum_q \int \Bigl [n_q\ln n_q+(1-n_q)
\ln (1-n_q) \Bigr ]d{\bf p},
\end{eqnarray}
which simplifies to
\begin{eqnarray}
{\cal S}(r)= \sum_q \Bigl [\frac {5}{3}J_{3/2}(\eta_q)/J_{1/2}(\eta_q)-\eta_q
\Bigr ]\rho_q.
\end{eqnarray}
The $n_q$'s in Eq.~(17)are the single-nucleon distributions; they are given by
\begin{eqnarray}
n_q(r,p,T)=\Bigl [1+\exp\{\frac {p^2}{2m_{q,k}T}-\eta_q\} \Bigr ]^{-1}.
\end{eqnarray}
The nucleon density $\rho_q(r)$ (given by Eq.~(14)) is obtained from
the momentum integration of the distribution function. Once the
entropy density is known, the free energy density ${\cal F}(r)$
of the system is obtained as
\begin{eqnarray}
{\cal F}(r)={\cal E}(r)-T {\cal S}(r).
\end{eqnarray}
For  an infinite system, the energy and free energy per nucleon  are then
calculated as $e={\cal E}/\rho $ and $f={\cal F}/\rho $. The symmetry
coefficient $e_{sym}(\rho,T)$ can then be computed from Eq.~(2). The 
coefficient $f_{sym}(\rho,T)$ can be similarly calculated.

\subsection{Modeling the hot nucleus}

The nucleon density profile for the hot nucleus is 
 computed self-consistently
in the finite temperature Thomas-Fermi (FTTF) approach. Calculations
in a box, as is usually done in the Thomas-Fermi procedure, however, lead 
to difficulties. 
At large distances from the
center, the density is small, the single-particle potential $V_q$ $\sim $0,
$\eta_q <<$0, and then $\rho_q \sim e^{\mu_q/T}$, a constant. 
At large distances from the nuclear center, the particle density then
does not vanish (at $T=$0, there are no problems though, $\mu_q$ being
negative, $\rho_q =$0 at large distance). The pressure of the system is
then nonzero, making the system thermodynamically unstable. The 
density profile also depends on the size 
of the box in which the FTTF calculations
are performed. The problem is overcome in the subtraction procedure
\cite{bonch,sur}, where the hot nucleus, assumed to be a thermalized system
in equilibrium with a surrounding gas representing the evaporated nucleons 
from the hot nucleus, is extracted 
from the embedding 
environment. This
method is based on the existence of two solutions to the FTTF
equations, one corresponding to the liquid phase with the surrounding
gas ($lg$) and the other corresponding to the gas phase ($g$) alone
\cite{sil}. The density of the thermalized nucleus in equilibrium
is given by $\rho_q=\rho_{lg}^q-\rho_g^q$. It is independent of the
box size, also goes to zero at large distances implying a vanishing
surface pressure. The nucleon number conservation gives 
\begin{eqnarray}
\int \Bigl [ \rho_{lg}^q(r)-\rho_g^q(r) \Bigr ] d^3r=N_q,
\end{eqnarray}
where $N_q$ is the number of neutrons or protons in the nucleus.
 The total energy $E$ of the nucleus is given by 
\begin{eqnarray}
E=E_{lg}-E_g,
\end{eqnarray}
where $E_{lg}$ and $E_g$ are the total energies of the liquid-gas system and
of the gas alone. From Eq.~(17), the total entropy 
of the nucleus 
can be recast as
\cite{des}, 
\begin{eqnarray}
 S=-\sum_q \int g_q(\epsilon_q,T)[f_q\ln f_q +(1-f_q)
\ln (1-f_q)]d\epsilon_q,
\end{eqnarray}
where $f_q$ is the single-particle occupation function in the energy
space,
\begin{eqnarray}
f_q(\epsilon_q,\mu_q,T)=\Bigl [ 1+\exp(\epsilon_q-\mu_q)/T \Bigr ]^{-1},
\end{eqnarray}
and $g_q$ is the subtracted single-particle level density.
It is given as \cite{sks1}
\begin{eqnarray}
g_q(\epsilon_q,T)&=&\frac {4\sqrt 2}{\pi \hbar^3} \int \Bigr [
(m_{q,k}^{lg})^{3/2}\sqrt {\epsilon_q-V_q^{lg}(r) } \nonumber \\
&&-(m_{q,k}^g)^{3/2} \sqrt {\epsilon_q-V_q^g(r) } \Bigr ] r^2dr.
\end{eqnarray}
The free energy is calculated from $F=E-TS$. In terms of the occupation
function, the density in the FTTF approximation 
can be written as
\begin{eqnarray}
\rho_q^i (r)&=&\frac {1}{2\pi^2\hbar^3}(2m_{q,k}^i(r))^{3/2} \nonumber \\
&&\times \int \sqrt {\epsilon_q-V_q^i(r) }f_q(\epsilon_q,\mu_q,T)d\epsilon_q.
\end{eqnarray}
Here $i$ stands for the liquid-gas phase or the gas phase.

\subsection{The energy mass}

The coupling of the nucleons to the dynamical surface phonons results
in an increased effective mass, with appearance of  
the so-called energy mass $m_{\omega }$,
as stated earlier. This effective mass decreases with temperature,
an increase in the symmetry energy is then anticipated. Taking the
energy mass into consideration, the total effective mass of the nucleon
$m^*$ can be written as 
\begin{eqnarray}
m^*=m(\frac {m_k}{m})(\frac {m_{\omega }}{m}),
\end{eqnarray}
where $m$ is the nucleon mass (for simplicity, we take the neutron-proton
mass difference to be zero as this is very small compared to
their masses). Aside from being  temperature-dependent, $m_\omega $ 
is density-dependent;
for a finite nucleus, its value therefore varies with 
position. A fully self-consistent calculation of $m_{\omega }$ is
thus very involved and not within the scope of the present work. We
therefore take a phenomenological form \cite{shlo} for $m_{\omega }$
such that
\begin{eqnarray}
\frac {m_{\omega }}{m} =1.0-0.4 A^{1/3}\exp \Bigl [ -
\bigl (\frac {T}{21A^{-1/3}} \bigr )^2
\Bigr ]\frac {1}{\rho (0)} \frac {d\rho (r)}{dr}.
\end{eqnarray}
The temperature $T$ and the distance $r$ are measured in MeV  and fm 
units, respectively,
$\rho (0)$ is the central density of the nucleon distribution
in the nucleus. The density in the above equation is $\rho (r)=
\rho_{lg}(r)-\rho_g(r)$; $A$ refers to the mass number of the subtracted
density (hereafter called the liquid mass  number). The collectivity implied
in $m_{\omega }$ refers to the liquid phase only, meaning thereby 
$m_{\omega }^{lg}=m_{\omega }$ and $m_\omega ^g=m$. 

The self-consistent calculation of the density profile with the
inclusion of $\omega$-mass is complex; a realistic extension
of the method given in Ref. \cite{shlo} is therefore adopted.
This is described in detail in Ref. \cite{des}.

The subtracted level density corresponding to Eq.~(25) is now
modified as
\begin{eqnarray}
\tilde g_q(\epsilon_q,T)&=&\frac {4\sqrt 2}{\pi \hbar^3}
\int \Bigl [ \bigl (m_{q,k}^{lg} \frac {m_\omega }{m} \bigr )^{3/2}
\sqrt {\epsilon_q-V_q^{lg}(r)\frac {m}{m_\omega }} \nonumber \\
&&-(m_{q,k}^g)^{3/2}\sqrt {\epsilon_q-V_q^g(r) }\Bigr ] r^2dr.
\end{eqnarray}
The densities in the $lg$ or $g$ phase modify accordingly
\begin{eqnarray}
\tilde \rho_q^i(r)&=&\frac {1}{2\pi^2\hbar^3}\Bigl [ 2m_{q,k}^i
\frac {m_\omega }{m} \Bigr ]^{3/2} \int \sqrt {\epsilon_q
-V_q^i \frac {m}{m_\omega ^i} } \nonumber \\
&&\times f_q(\epsilon_q,\tilde \mu_q,T) d\epsilon_q .
\end{eqnarray}
 The chemical potential $\mu_q$ modifies to $\tilde \mu_q$ to
conserve the particle number in the nucleus
\begin{eqnarray}
N_q=\int \tilde g_q(\epsilon_q,T)f_q(\epsilon_q,\tilde \mu_q, T)d\epsilon_q.
\end{eqnarray}
The energy and the free energy of the nucleus is calculated with the modified
density given by Eq.~(30).

\subsection{The symmetry coefficients}

For homogeneous infinite matter, we have calculated the symmetry energy
coefficients using Eq.~(7). The symmetry free energy coefficients are
likewise calculated,
\begin{eqnarray}
f_{sym}(\rho,T)=f(\rho,\delta=1,T)-f(\rho,\delta=0,T).
\end{eqnarray}
The kinetic $(K)$ and potential $(I)$ components of the symmetry coefficients
$e_{sym}^K, e_{sym}^I, f_{sym}^K$ and $f_{sym}^I$ are similarly
calculated replacing the energy $e$ and free energy $f$ (in Eq.~(7) and
(32)) by $e_K$ or $e_I$ and $f_K$ or $f_I$, respectively where $e_K$,
$e_I$, say, stand for the kinetic and interaction parts of the energy
per nucleon.

For  a  finite nucleus (with Coulomb interaction turned on),
the symmetry energy coefficient 
in the difference method is defined as
\begin{eqnarray}
e_{sym}(A,T)= \Bigl [e_n(A,I_1,T)-e_n(A,I_2,T) \Bigr ]/(I_1^2-I_2^2),
\end{eqnarray}
where  the $e_n$'s are the nuclear part of the energy per nucleon of the
nuclear pair of mass $A$ having different isospins $I_1$ and $I_2$.
In the local density approximation (LDA), one 
defines the symmetry 
energy coefficient for a specific nucleus once its density profile
is known. It is given as
\begin{eqnarray}
e_{sym}(A,T)=\frac {1}{I^2A}\int \rho (r) e_{sym}[\rho (r),T]
\delta_l^2(r) d^3r
\end{eqnarray}
In Eq.~(34), $e_{sym}[\rho (r),T]$ is the symmetry coefficient 
at temperature $T$ of infinite matter at the value of the local
density $\rho (r)$, $\rho_n(r)$ and $\rho_p(r)$ are the neutron
and proton densities and 
$\delta_l (r)$ is the isospin asymmetry of the
local density. The symmetry free energy coefficient can be defined exactly
in a parallel way in the difference method and also in LDA.

As already discussed, both the difference method and the LDA fail to give 
unique values for the symmetry energies for finite nuclei or their
temperature dependence. A possible means to arrive at unambiguous
values of the temperature-dependent symmetry coefficients can be
achieved in the framework of the liquid-drop model, exploiting
the Bethe-Weizs\"acker mass formula. One 
can calculate  the energies
and free energies of a set of nuclei spanning nearly the whole
periodic table as a function of temperature in the subtracted 
FTTF procedure with effects from energy-mass taken into account
and expand the energy or free energy in terms of macroscopic 
parameters as 
\begin{eqnarray} 
E(N,Z,T)&=&a_v(T)A+a_s(T)A^{2/3}+a_c\frac {Z^2}{A^{1/3}}  \nonumber \\
&&+e_{sym}(A,T)I^2A,
\end{eqnarray} 
\begin{eqnarray} 
F(N,Z,T)&=&f_v(T)A+f_s(T)A^{2/3}+a_c\frac {Z^2}{A^{1/3}}  \nonumber \\
&&+f_{sym}(A,T)I^2A.
\end{eqnarray} 
The coefficients $a_v,a_s$ and $a_c$  are the volume, surface and
Coulomb energy coefficients; 
similarly $f_v$ and $f_s$ refer to the volume and
surface free energies. The Coulomb energy and free energy are the same. 
Nuclei are finite systems, they have varying density profiles. This
necessitates introduction of a mass-dependent surface component
in $e_{sym}(A,T)$ over and above the mass-independent volume component
$e_{sym}^v(A,T)$. Two definitions have been used to incorporate the
mass dependence in $e_{sym}(A,T)$. The first, hereafter referred to
as I \cite{mye2,dan1,dan2} is 
\begin{eqnarray} 
e_{sym}(A,T)=\frac {e_{sym}^v(T)}{1+\frac {e_{sym}^v(T)}{\beta_E(T)}A^{-1/3}},
\end{eqnarray} 
and the second, referred to as II \cite{lip,rei,jia} is,
\begin{eqnarray} 
e_{sym}(A,T)= e_{sym}^v(T)-e_{sym}^s(T)A^{-1/3}.
\end{eqnarray} 
In Eq.~(38), the first term on the right hand side (rhs) measures
the contribution from the nearly constant value of the density
from the central part of the nucleus; at $T=0$, $e_{sym}^v$ is to be equated
to $e_{sym}$ corresponding to infinite nuclear matter at saturation
density. The second term  in Eq.~(38) is the surface
symmetry energy, accounting for the contribution coming from the
surface profile specific to the nuclear mass number. In Eq.~(37),
$\beta_E(T)$ is a measure of the surface symmetry energy. In the 
limit of large $A$, $(e_{sym}^v(T))^2/{\beta_E(T)} \sim e_{sym}^s(T)$.
The phenomenological value of $e_{sym}^s(T=0)$ is $\sim $ 50-58 MeV
\cite{jia,sto,rei} and that of $e_{sym}^v(T=0)/\beta_E(T=0)$ is
in the close range of $\sim $ 2.4 $\pm $0.4 \cite{dan1,dan2,liu}.

Since the Coulomb energies are precisely known in a FTTF calculation,
one can make a four-parameter fit of only the nuclear part of the
energies $E_n$ and free energies $F_n$ of nuclei,
\begin{eqnarray} 
E_n(N,Z,T)=a_v(T)A+a_s(T)A^{2/3}+e_{sym}(A,T)I^2A,
\end{eqnarray} 
\begin{eqnarray} 
F_n(N,Z,T)=f_v(T)A+f_s(T)A^{2/3}+f_{sym}(A,T)I^2A.
\end{eqnarray} 
Here, $e_{sym}(A,T)$ is given by Eq.~(37) or (38). Similarly, 
$f_{sym}(A,T)$ is defined as,
\begin{eqnarray} 
f_{sym}(A,T)=\frac {f_{sym}^v(T)}{1+\frac {f_{sym}^v(T)}{\beta_F(T)}A^{-1/3}},
\end{eqnarray} 
\begin{eqnarray} 
f_{sym}(A,T)= f_{sym}^v(T)-f_{sym}^s(T)A^{-1/3}.
\end{eqnarray} 
The four-parameter set $f_v, f_s, f_{sym}^v$ and $f_{sym}^s$ (or $\beta_F$)
has the same connotation as the set $a_v, a_s, e_{sym}^v$ and 
$e_{sym}^s$ (or $\beta_E$), except that the former set refers to
free energy.

\section{Results and Discussions}

 To study the temperature dependence of the symmetry coefficients of
infinite and finite nuclear systems, we have used three interactions,
all from the Skyrme class, namely SkM*, SLy4 and SK255. These interactions
describe the ground state energies of finite nuclei rather well, 
there is some difference though in the computed 
values of some of the macroscopic
observables as shown in Table I.

 \begin{figure}[ht]
\resizebox{3.0in}{!}{ \includegraphics[]{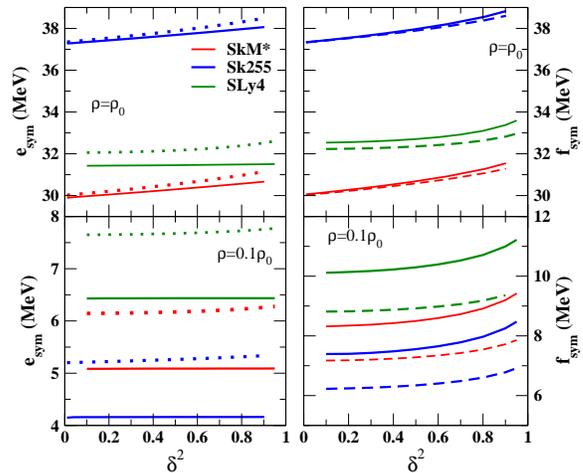}}
\caption{\label{fig1:snm1} (Color online) The symmetry energy coefficient
$e_{sym}$ and symmetry free energy coefficient $f_{sym}$ of infinite 
nuclear matter as defined from Eq.~(6) shown as a function of the asymmetry
parameter  $\delta^2$ at two densities $\rho =\rho_0$ and at 
$\rho $= 0.1 $\rho_0$ for three interactions, SkM$^*$ (red),
SK255 (blue) and SLy4 (green). The dotted lines refer to
calculations at $T =$ 0.0 MeV, the dashed lines at $T =$ 5.0 MeV and the 
full  lines at $T =$ 8.0 MeV, respectively.} 
  \end{figure}

 The symmetry coefficients $e_{sym}$ and $f_{sym}$ as calculated from
Eq.~(6) for {\it infinite matter} are displayed 
in Fig.~1 as a function of the isospin
asymmetry $\delta^2$ for the three interactions at the corresponding
saturation densities $\rho_0$ and at one-tenth of this density.  The left
panels refer to $e_{sym}$, the right panels to $f_{sym}$. The dotted lines
represent calculations at $T$ = 0, the dashed lines at $T$ = 5 and full
lines at $T$ = 8 MeV, respectively.  A non-zero slope at all densities
and temperatures  for most of the lines shows that the symmetry energy
or symmetry free energy content in the system is nonlinear in $\delta^2$,
but the small value of the slope (as seen from the slow rise of $e_{sym}$
or $f_{sym}$ with $\delta^2$) 
shows  that the nonlinearity is weak. 
From now on, we use the definition as given by
Eq.~(7) for the symmetry coefficients for infinite nuclear matter.

The temperature dependence of the symmetry energy coefficients 
corresponding to  the three interactions is displayed in Fig.~2. The upper 
panels pertain to the saturation density $\rho_0$, the lower
panels to $\rho = \rho_0/10 $. At the saturation density, it is
seen that $e_{sym}$, shown by the full lines (whose values are  somewhat
different for different interactions), has a very weak temperature
dependence. 
The behavior for the interaction component
$e_{sym}^I$ (the dashed lines) and the kinetic energy component
$e_{sym}^K$ (shown by the dotted lines) are similar. 
It may be mentioned that the kinetic energy component is evaluated
with the bare nucleon mass.
At the sub-saturation 
density, $e_{sym}$ shows a slow fall \cite{xu,mou,de4}.
This is essentially due to the decrease in its kinetic energy part
with temperature. The lesser importance of  the Pauli blocking because of
the increased diffuseness of the nucleon Fermi surfaces with rising temperature 
lies behind this decrement in $e_{sym}^K$ at lower densities.
At these densities, the symmetry free energy coefficient displays,
on the contrary, a comparatively more prominent rise as seen 
from the lower panel of Fig.~3. This is in fair agreement with
those obtained earlier \cite{xu,de4}. The increase in $f_{sym}$
with temperature at very low densities can be understood from 
the fact that $s_{sym}$, the symmetry entropy coefficient is then
negative, $\sim $ -$\frac {1}{2}$ \cite{de6}. Thus, even if $e_{sym}$
falls slower, the rise in $f_{sym}$ is noticeable at these densities. 
At higher density, $f_{sym}$ as with $e_{sym}$ shows nearly no
trace of temperature dependence; this is shown 
in the upper panel of Fig.~3 for $\rho = \rho_0$.

 \begin{figure}
\resizebox{3.0in}{!}{ \includegraphics[]{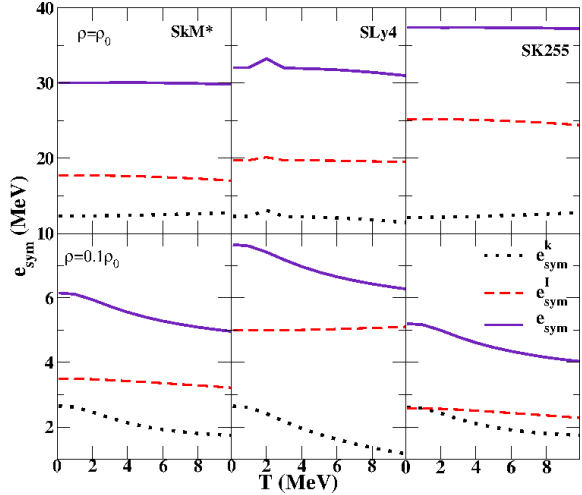}}
\caption{\label{fig2:snm1} (Color online) The thermal evolution 
of the symmetry energy coefficients $e_{sym}$ 
as defined in Eq.(7) and its interaction
component $e_{sym}^I$ and kinetic energy component $e_{sym}^K$
shown at densities $\rho =\rho_0$ and at $\rho $=0.1$\rho_0$
for the three interactions as shown. The full violet line corresponds to
$e_{sym}$, the black dotted line refers to $e_{sym}^K$ and the red
dashed line to $e_{sym}^I$. } 
  \end{figure}

In the previous section, we have already adumbrated that the symmetry
energy coefficients of {\it finite nuclei} may not come out unique in the LDA
(Eq.~(34)) or in the difference method (Eq.~(33)). As a demonstration,
we show this in Figs.~4$-$6 
where the thermal evolution of the symmetry
coefficients 
has been calculated in LDA and in the difference method
with the SLy4 interaction. In the panels of Fig.~4, the temperature
dependence of $e_{sym}$, calculated in LDA  for $A$ = 56, 150 and 208
is displayed.  The values of $e_{sym}$ decrease with temperature, but it is
evident that for a given isobar, the symmetry energy coefficient depends
on the choice of the $(Z,N)$ value, the isobar with the higher atomic number
having larger symmetry energy. Figs.~5 and 6 show $e_{sym}$ and $f_{sym}$
evaluated in the difference method for the nuclei $A$ = 56 and $A$ =
112 taken  as representative examples.  The results again are seen to
depend on the choice of the nuclear pairs (shown by the proton numbers in
brackets) for the isobars. The decrease
in $e_{sym}$ (Fig.~4 and upper panels of Figs.~5 and 6) or the increase
in $f_{sym}$ (lower panels  of  Figs.~5 and 6) with temperature for finite
nuclei comes from the increased weight from the temperature-dependent
surface profile of lower density.
 \begin{figure}
\resizebox{2.0in}{!}{ \includegraphics[]{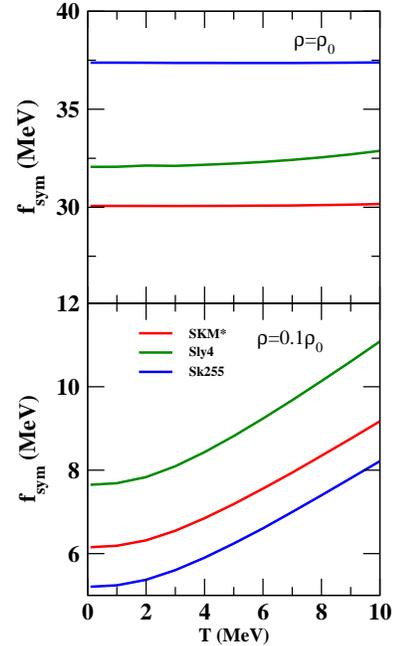}}
\caption{\label{fig3:snm1} (Color online) The temperature dependence 
of the symmetry free energy coefficient  of infinite nuclear matter
at two densities $\rho =\rho_0$ and at $\rho = 0.1\rho_0$ as shown
for the three interactions, SkM$^*$ (red line), SLy4 (green line)
and SK255 (blue line).} 
  \end{figure}

As already stated, unique values of symmetry 
coefficients could be ascribed to finite nuclei and their temperature
dependence obtained, albeit in an average sense, 
making use of the  Bethe-Weizs\"acker
mass formula. To this end, a set of spherical and near-spherical 
nuclei covering almost the entire periodic table is chosen.
We take 69 nuclei, having masses as 36 $\le  A \le $ 218 and 
atomic charges 14 $\le Z \le $ 92
(the list of nuclei is taken from Ref. \cite{klu}). Their energies
and free energies are then calculated in a temperature grid in the 
subtracted FTTF procedure, taking into account the dressing of the 
nucleon mass to energy mass $m_\omega $ arising from the coupling 
of the nucleonic motion to the surface vibrations. The nuclear part 
of the energies and free energies (i.e., the Coulomb energy is 
subtracted from the total energies or free energies) 
are then fitted with a four-parameter
fit. The parameter sets are ($a_v, a_s, e_{sym}^v$ and $\beta_E$) or
($f_v, f_s, f_{sym}^v$ and $\beta_F$) for energies or free energies 
in definition I. In definition II, the parameters are ($a_v, a_s, e_{sym}^v$
and $e_{sym}^s$) or ($f_v,f_s, f_{sym}^v$ and $f_{sym}^s$).
We leave here discussions on the temperature dependence of
$a_v, f_v$ or $a_s, f_s$.  They were discussed, for a different
set of nuclear interactions in Ref. \cite{de5}. For the three
Skyrme-class interactions used here, $a_v(T), f_v(T), a_s(T)$
and $f_s(T)$ follow nearly the same pattern.

 \begin{figure}
\resizebox{2.0in}{!}{ \includegraphics[]{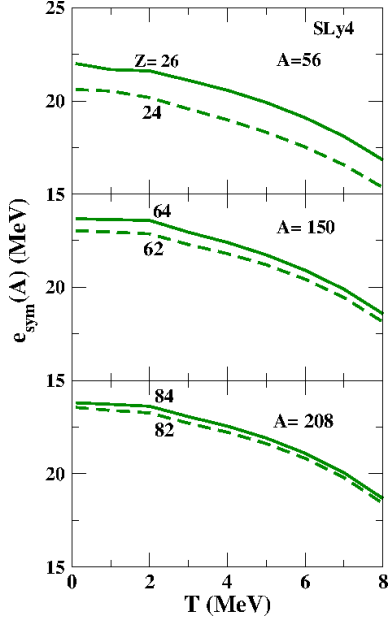}}
\caption{\label{fig4:snm1} (Color online) The temperature dependence of the
symmetry coefficient $e_{sym}(A)$ calculated in the local density
approximation (LDA, Eq.~(34)) with the SLy4 interaction is shown for three
nuclei with mass numbers $A =$ 56, 150, and 208.
The full lines correspond to calculations for isobars with the higher
proton numbers, the dashed lines refer to those with 
proton numbers less by two.} 
  \end{figure}

 \begin{figure}
\resizebox{2.0in}{!}{ \includegraphics[]{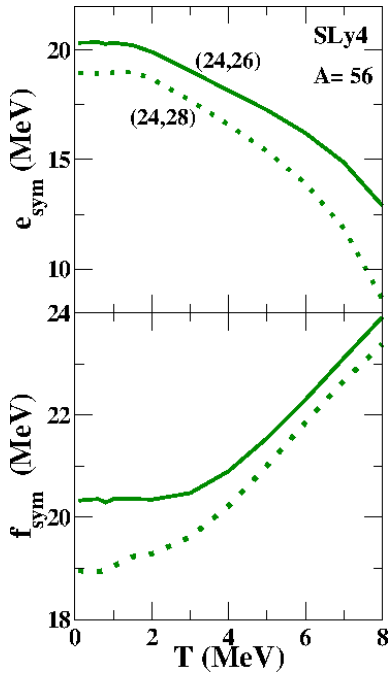}}
\caption{\label{fig5:snm1} (Color online) The temperature dependence
of the symmetry coefficients $e_{sym}$ and $f_{sym}$ calculated
in the difference method (Eq.(33)), shown for the nucleus $A =$ 56 with
the SLy4 interaction. The full and dashed lines correspond  to
the isobars with different charge pairs.
The charge numbers for the nuclear pair are given in the bracket.}
  \end{figure}

 \begin{figure}
\resizebox{2.0in}{!}{ \includegraphics[]{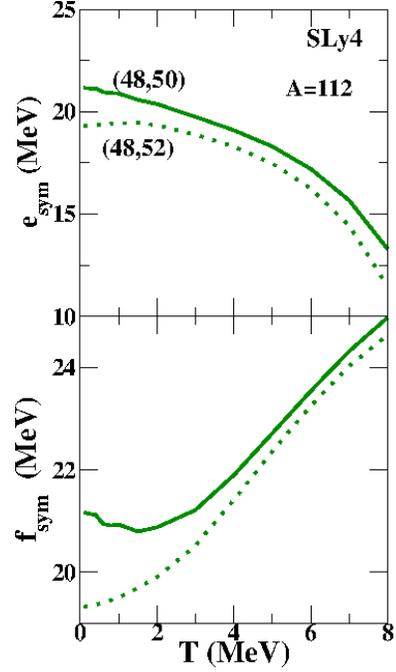}}
\caption{\label{fig6:snm1} (Color online) The same as in Fig.~5
for the nucleus with mass number $A =$ 112.} 
  \end{figure}
In Fig.~7, the thermal evolution of the 
volume part of the symmetry energy coefficient 
$e_{sym}^v$ (upper panels) and symmetry free energy coefficient $f_{sym}^v$
(lower panels) 
for the three interactions 
is shown. The left panels refer to 
definition I, the right panels to definition II. The behavior 
of $e_{sym}^v$ depends on how $e_{sym}(A)$ is defined. In definition
I, it generally falls with temperature (for SK255 interaction,
the fall is slow though). In definition II, a slow increase 
in noticeable. The coefficient $f_{sym}^v$, on the contrary,
shows a rise with temperature in definition I; in definition II,
it is nearly temperature-independent.

 \begin{figure}
\resizebox{3.0in}{!}{ \includegraphics[]{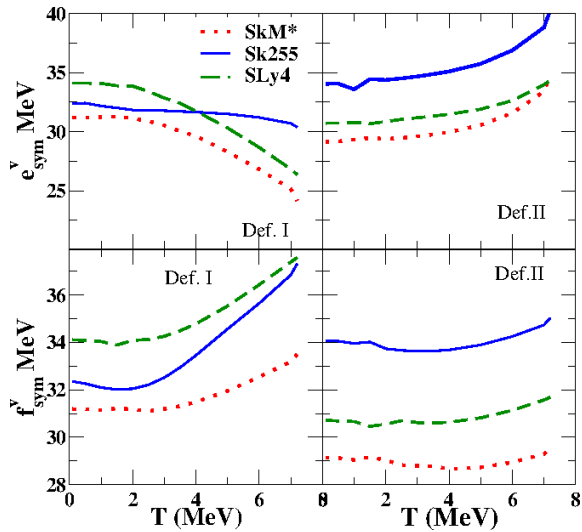}}
\caption{\label{fig7:snm1} (Color online) The volume symmetry 
energy coefficient $e_{sym}^v$ and symmetry free energy coefficient
$f_{sym}^v$, shown as a function of temperature for three interactions
as displayed. The left panels present results 
calculated  with the definition I (Eq.~(37)), the right 
panels do so with the definition II (Eq.~(38)).} 
\end{figure}
The thermal dependence of the coefficients $\beta_E$ or $\beta_F$ 
for the three interactions is shown in the upper and lower panels of
Fig.~8. At $T$ = 0, the value of $\beta_E$ or $\beta_F$ lies between
$\sim $ 12-13 MeV for the three interactions, in remarkable consonance
with the empirical value of $\sim $ 13 MeV obtained from the analyses of
the 'experimental' symmetry energies of isobaric nuclei \cite{liu}. With
temperature, $\beta_E$ decreases (for SK255 interaction, the decrease
is again very slow),  while $\beta_F$ generally increases. It is,
however, noticed that both $e_{sym}^v/\beta_E$ and $f_{sym}^v/\beta_F$
are nearly temperature independent and also interaction independent
in our calculations. They lie in the  vicinity of $\sim $ 2.63 $\pm $
0.01; this is very close to $\sim $ 2.64 $\pm $ 0.01 obtained earlier
with the SBM and KDE0 interactions \cite{de5}.

The temperature dependent surface symmetry coefficients $e_{sym}^s$
and $f_{sym}^s$ 
calculated using Eq.~(38) in definition II  are displayed 
in the upper and lower panels of Fig.~9. At $T$ = 0, $e_{sym}^s$
(also $f_{sym}^s$) is $\sim $ 41 MeV for the SkM$^*$ and SLy4 interactions,
close to the value of $\sim $ 45 MeV obtained by Stoitsov et al. 
\cite{sto}, whereas for the SK255 interaction it is $\sim $ 58 MeV,
nearly equal to that obtained from the binding
energy analysis in Ref. \cite{jia}. As the temperature increases,
$e_{sym}^s$  increases sharply pointing to the growing importance
of the surface in the calculation of $e_{sym}(A)$. In comparison,
$f_{sym}^s$ shows a very slow decline with temperature for all
the three interactions. Such a behavior was also noted in our
earlier calculations with the SBM and KDE0 interactions \cite{de5}.
A comparison of our results with those in Ref. \cite{lee} reveals that
in both calculations, the surface symmetry coefficients are more 
temperature sensitive than the volume symmetry coefficients.  However,
in the calculations of \cite{lee}, the temperature dependence of surface 
coefficients are comparatively  more pronounced than those of ours. 
There are other subtle differences too;
the density profiles used in \cite{lee} are not self-consistent.
Also,  high-density approximation in the
whole temperature domain was used there.

 \begin{figure}
\resizebox{2.0in}{!}{ \includegraphics[]{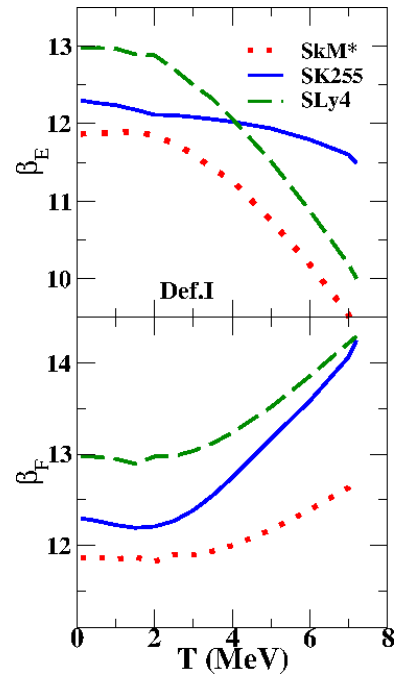}}
\caption{\label{fig8:snm1} (Color online) The thermal dependence
of the coefficients $\beta_E$ and $\beta_F$ 
(from definition I,  Eq.~(37))
for the three interactions are shown.} 
  \end{figure}

 \begin{figure}
\resizebox{2.0in}{!}{ \includegraphics[]{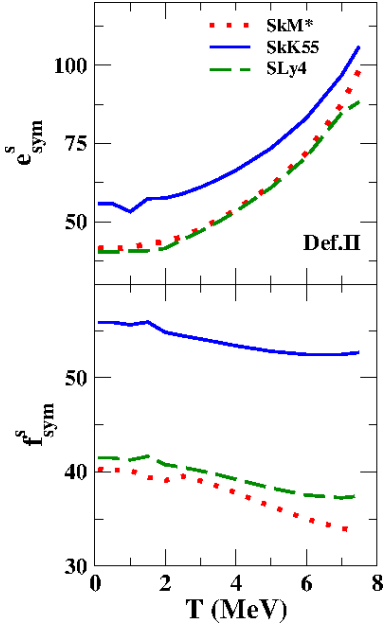}}
\caption{\label{fig9:snm1} (Color online) The thermal dependence
of the surface symmetry coefficients $e_{sym}^s$ and $f_{sym}^s$
(from definition II, Eq.~(38)) displayed for the three interactions.} 
  \end{figure}

 \begin{figure}
\resizebox{3.0in}{!}{ \includegraphics[]{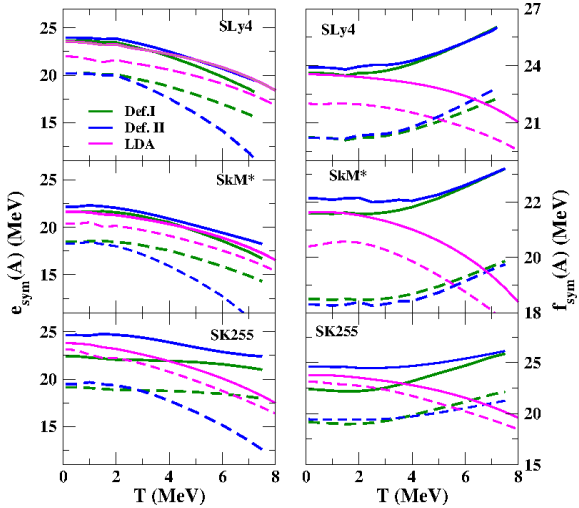}}
\caption{\label{fig10:snm1} (Color online) The temperature dependence
of the symmetry coefficients of finite nuclei shown for the three
interactions in definition I (Eq.~(37), green line), 
definition II (Eq.~(38), blue line) and 
in LDA (magenta line). The dashed lines correspond to calculations
for $A =$56, the full lines refer to those for $A =$ 208. The 
left panels display results for $e_{sym}(A)$, the right panels
do so for $f_{sym}(A)$. For more details, see text.}
  \end{figure}

 \begin{figure}
\resizebox{3.0in}{!}{ \includegraphics[]{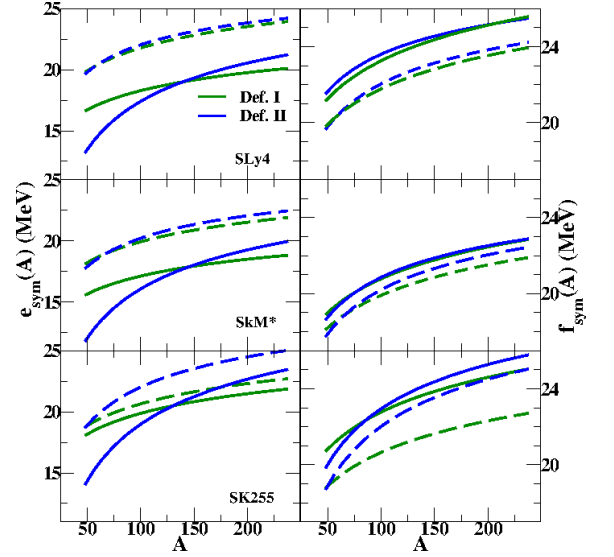}}
\caption{\label{fig11:snm1} (Color online) The mass number  dependence of the
symmetry coefficients $e_{sym}(A)$ (left panels) and $f_{sym}(A)$
(right panels) at two temperatures calculated with the three interactions.
The green lines refer to def. I, the blue lines to def. II. The
dashed lines correspond to results at $T =$0.0, the full lines
do so at $T =$ 6.0.} 
  \end{figure}
Once the volume and surface symmetry coefficients are obtained as 
a function of temperature, the temperature-dependent symmetry coefficients
for finite nuclei can be obtained unambiguously using definition I
(Eqs.~(37) and (41)) or definition II (Eqs.~(38) and (42)) for $e_{sym}(A)$
and $f_{sym}(A)$. In Fig.~10, for two representative nuclei with
$A$ = 56 and $A$ = 208, $e_{sym}(A)$ and $f_{sym}(A)$ are displayed
as a function of temperature in the left and right panels, respectively,
for the three interactions with the volume and surface symmetry coefficients
so obtained. 
The full lines refer to $A$ = 208, the broken lines to $A$
= 56. The green color corresponds to definition I, the blue color
to definition II. The general findings
are: for all the three interactions, $e_{sym}(A)$ decreases with
temperature, $f_{sym}(A)$ remains nearly unchanged upto $T \sim $ 3 MeV,
and then increases slowly. To have a feel about the applicability
of the local density approximation in calculating the thermal
behaviour of the symmetry coefficients, results in LDA for
$^{56}$Fe and $^{208}$Pb, the most stable nuclei corresponding 
to their mass numbers are also shown (in magenta color). 
In LDA, both $e_{sym}(A)$ and $f_{sym}(A)$ 
are seen to register a fall with increasing
temperature.

The mass dependence of $e_{sym}(A)$ and $f_{sym}(A)$, calculated
in  definitions I and II is displayed in the left and right panels of 
Fig.~11 for the interactions at two temperatures, $T$ = 0.0 and $T$
= 6.0 MeV. The green color again corresponds to definition I,
the blue color refers to definition II.  The dashed lines represent results 
at $T$ = 0.0, the full lines do so at $ T$ = 6.0 MeV. The general
observations from 
Fig. 11 are that at fixed temperature, both
$e_{sym}(A)$ and $f_{sym}(A)$ increase with $A$; this follows from
the definition. It is also found that at a fixed mass number, 
$e_{sym}(A)$ decreases when temperature is raised whereas $f_{sym}(A)$
increases. These observations are true for both the definitions
of the symmetry coefficients and collate with the 
already obtained understanding from Figs.~7, 8 and 9 
on the thermal 
evolution of $f_{sym}^v$, $f_{sym}^s$, and $\beta_F$.

\section{Concluding remarks}
 
Results from calculations on the temperature dependence of 
symmetry energy and symmetry free energy coefficients of
infinite nuclear matter and also of finite nuclei, done in
a finite-temperature-Thomas-Fermi model, are reported in this
paper. Three Skyrme-class interactions are chosen, namely, SkM$^*$,
SLy4 and SK255. For infinite matter, we have made investigations
at different densities, near saturation and also at densities in the
sub-saturation region. Near saturation density, both the symmetry energy
and symmetry free energy coefficients $e_{sym}$ and $f_{sym}$ show a very
weak temperature dependence.  At lower densities, however, $e_{sym}$
decreases with temperature whereas $f_{sym} $ displays a comparatively
more prominent rise.  For finite systems, in order to give stability to
the evaporating nuclei, calculations have been done in the 
thermal Thomas-Fermi
framework with subtraction. The Thomas-Fermi calculations
are static in nature.  Dynamical effects from the coupling of the surface
phonons to the intrinsic particle motion are taken into account through
a phenomenological parametric form of the nucleon energy mass $m_\omega
$. At relatively low temperatures, the $\omega $-mass lends a heaviness
to the nucleon effective mass; this has an appreciable effect on the
symmetry coefficients. Once effects due to $\omega $-mass are taken
into account, the symmetry coefficients are seen to decrease somewhat,
particularly at low temperatures.
\begin{table}[t]
\begin{center}
\caption{Nuclear matter properties for the SkM*, SLy4 and SK255
interactions considered in this work.}
\begin{tabular}{ccccccc}
\hline
   Force&$e(\rho_0,0,0)$&$\rho_0$  & $ K_v$ & $  m^*/m$  &$e_{sym}$   & $  L$  \\
&(MeV) &  ($fm^{-3}$)  & (MeV)& &(MeV)&(MeV)          \\
\hline
 SK255& 16.3& 0.157&  254.9& 0.80& 37.4&  95 \\
 SkM* & 15.8& 0.160&  216.6& 0.79& 30.0&  46 \\
 SLy4 & 16.0& 0.160&  229.9& 0.70& 32.0&  46 \\
\hline
\end{tabular}
\label{data-gqr}
\end{center}
\end{table}

\begin{table}[t]
\begin{center}
\caption{The values of the Skyrme parameters for SkM*, SLy4 and SK255
interactions.}

\begin{ruledtabular}
\begin{tabular}{lddd}
\multicolumn{1}{c}{Parameters}& 
\multicolumn{1}{c}{SkM*}&
\multicolumn{1}{c}{SLy4}&
\multicolumn{1}{c}{SK255}\\
\hline
$t_0$(MeV fm$^3$)&-2645.0&-2488.91&-1689.35 \\
$t_1$(MeV fm$^5$)&410.0&486.82&389.30  \\
$t_2$(MeV fm$^5$)&-135.0&-546.39&-126.07 \\
$t_3$(MeV fm$^{3(\alpha +1)}$)&15595.0&13777.0&10989.59 \\
$x_0$&0.09&0.834&-0.1461 \\
$x_1$&0.0&-0.344&0.116 \\
$x_2$&0.0&-1.0&0.0012 \\
$x_3$&0.0&1.354&-0.7449 \\
$\alpha $&0.1666&0.1666&0.3563 \\
\end{tabular}
\end{ruledtabular}
\label{data-gqr}
\end{center}
\end{table}

In understanding the symmetry coefficients $e_{sym}(A)$ and
$f_{sym}(A)$ for finite nuclear systems and their thermal evolution,
some ambiguities about their proper definition could be noted. 
We have explored the usual difference method \cite{dea}, also the
one obtained from the LDA \cite{sam1}. In both these definitions,
further specifications other than the mass number $A$ of the nucleus are
necessary. In both these definitions, $e_{sym}(A)$ decreases
with temperature. The symmetry free energy coefficient also
falls with temperature in LDA, however, an opposite trend is 
observed in the difference method.

The ambiguities arising from these definitions of the symmetry
coefficients of atomic nuclei are resolved from calculations
of the symmetry energies and symmetry free energies of a host of nuclei
along the periodic table as a function of temperature in the 
microscopic FTTF model with subtraction. The modification to the 
energies and free energies from the temperature-dependent energy
mass are taken into account. The temperature dependence of the 
symmetry coefficients  is then obtained from a liquid-drop-inspired
fit of the total energies and free energies of these systems of 
nuclei. The two components, volume ($e_{sym}^v,f_{sym}^v$) and
surface ($e_{sym}^s,f_{sym}^s$) that make up the total symmetry
coefficients of finite systems have a temperature dependence
that is nearly independent of the energy functionals chosen to
calculate their energies. The volume symmetry energy coefficient
$e_{sym}^v$ shows a strong temperature dependence, the surface
part $e_{sym}^s$ displays an even stronger sensitivity.
This results in a rapid fall of $e_{sym}(A)$ as the temperature rises.
The thermal sensitivity of the symmetry free energy coefficients 
is  comparatively weaker. These calculations, in addition, give
information on the thermal evolution of the volume and surface 
energies of nuclei. They are in excellent agreement with those
in common usage, but because of the different focus in this
communication are not reported here.

\begin{acknowledgments}
The authors gratefully acknowledge the assistance of Tanuja Agrawal
in the preparation of the manuscript.
J.N.D  acknowledges the support of DST, Government of India.
M.C. and X.V. acknowledge the support of the Consolider Ingenio 2010
Programme CPAN CSD2007-00042, of the grants 
FIS2011-24154 from MICINN and FEDER, and of grant 2009SGR-1289 from
Generalitat de Catalunya. 
\end{acknowledgments}

\end{document}